\documentclass[12pt]{article}
 \pdfoutput=1
\usepackage{amssymb}
\usepackage{amsmath}
\usepackage{amstext}
\usepackage{graphicx,epsfig}
\usepackage{epsfig}
\usepackage{verbatim} 
\usepackage{fancybox}
\usepackage{color}
\usepackage{ulem}
\usepackage{enumitem}
\usepackage{subfigure}
\usepackage{bbm}
\usepackage{parskip}
\usepackage{cite}
\usepackage{youngtab}

\newcommand{\Comment}[1]{{}}
\definecolor{MyDarkBlue}{rgb}{0.15,0.15,0.45}
\usepackage[linktocpage=true]{hyperref}
\hypersetup{
colorlinks=true,
citecolor=MyDarkBlue,
linkcolor=MyDarkBlue,
urlcolor=MyDarkBlue,
pdfauthor={Kurt Hinterbichler},
pdftitle={},
pdfsubject={hep-th}
}

\newcommand{\be}{\begin{equation}}
\newcommand{\ee}{\end{equation}}
\newcommand{\bea}{\begin{eqnarray}}
\newcommand{\eea}{\end{eqnarray}}
\newcommand{\beas}{\begin{eqnarray*}}
\newcommand{\eeas}{\end{eqnarray*}}
\newcommand{\nn}{\nonumber}

\def\({\left(}
\def\){\right)}

\newcommand{\Tr}{\text{Tr}}

\newcommand{\half}{\frac{1}{2}}

\numberwithin{equation}{section}

\begin{document}


\begin{center}
{\LARGE { Manifest Duality Invariance for the  \\ \vspace{.2cm}  Partially Massless Graviton }}
\end{center} 
 \vspace{1truecm}
\thispagestyle{empty} \centerline{
{\large  { Kurt Hinterbichler}}\footnote{E-mail: \Comment{\href{mailto:khinterbichler@perimeterinstitute.ca}}{\tt khinterbichler@perimeterinstitute.ca}}
                                                          }

\vspace{1cm}

\centerline{{\it 
Perimeter Institute for Theoretical Physics,}}
 \centerline{{\it 31 Caroline St. N, Waterloo, Ontario, Canada, N2L 2Y5 }} 

\begin{abstract}

For a special value of the mass, a massive graviton on de Sitter space acquires an enhanced scalar gauge symmetry, and is called partially massless.
The partially massless graviton possesses a duality invariance akin to electromagnetic duality.  We display this duality in its manifestly local and covariant form, in which it acts to interchange the first order field equations and Bianchi identities of a gauge invariant field strength.

\end{abstract}



\section{Introduction}
\parskip=5pt
\vspace{1cm}

The action for a free massive graviton of mass $m$, carried by a symmetric tensor $h_{\mu\nu}$, propagating on a background $3+1$ dimensional de Sitter space $g_{\mu\nu}$ with cosmological constant $\Lambda$, and coupled to a source $T_{\mu\nu}$ is
 \bea \nn S&=&\int d^4x\ \sqrt{-g}\left[ -{1\over 2}\nabla_\lambda h_{\mu\nu} \nabla^\lambda h^{\mu\nu}+\nabla_\lambda h_{\mu\nu} \nabla^\nu h^{\mu\lambda}-\nabla_\mu h\nabla_\nu h^{\mu\nu}+\half \nabla_\mu h\nabla^\mu h\right. \\ &&\left. +\Lambda\left( h^{\mu\nu}h_{\mu\nu}-\half h^2\right)-\frac{1}{2}m^2(h_{\mu\nu}h^{\mu\nu}-h^2)+h_{\mu\nu}T^{\mu\nu}\right]. \label{curvedmassivelin}\eea
When the graviton mass takes the special value
\be m^2={2\Lambda\over 3}\, ,\label{masstuning}\ee
the theory has an enhanced gauge symmetry 
\be \delta h_{\mu\nu}=\nabla_\mu\nabla_\nu\, \phi+{\Lambda\over 3} g_{\mu\nu}\,\phi\, ,\label{gaugesym}\ee
with a scalar gauge parameter $\phi$, provided the source satisfies the conservation condition
\be \nabla_\mu\nabla_\nu T^{\mu\nu}+{\Lambda\over 3} T^\mu_{\ \mu}=0\, .\label{Tconservat}\ee
A massive graviton with the special value \eqref{masstuning} is known as a partially massless graviton \cite{Deser:1983mm,Higuchi:1986py,Brink:2000ag,Deser:2001pe,Deser:2001us,Deser:2001wx,Deser:2001xr,Zinoviev:2001dt}.  It propagates 4 healthy/unitary degrees of freedom, one fewer than the 5 polarizations of a generic massive spin-2, because the longitudinal polarization is eliminated by the gauge invariance \eqref{gaugesym}.

The partially massless theory is of interest as a gravitational theory because the relation \eqref{masstuning}, enforced by the gauge symmetry \eqref{gaugesym}, ties the value of the cosmological constant to the value of the graviton mass.  Since the graviton mass can itself be naturally small due to the enhanced diffeomorphism invariance of general relativity at $m=0$ \cite{deRham:2012ew,deRham:2013qqa}, this offers an attractive avenue towards solving the cosmological constant problem.  Unfortunately, there are obstructions to realizing a fully non-linear theory with these properties \cite{Zinoviev:2006im,Deser:2012qg,deRham:2013wv,Joung:2014aba,Zinoviev:2014zka,Garcia-Saenz:2014cwa}.

The partially massless theory possesses many features that are more akin to photons than to gravitons, including conformal invariance and null propagation in four dimensions \cite{Deser:2004ji}, a scalar gauge invariance \eqref{gaugesym}, the existence of a one-derivative gauge invariant field strength \cite{Deser:2006zx}, and the aforementioned difficulties with non-trivial self-interactions.  Another photon-like feature recently uncovered in \cite{Deser:2013xb} will be our focus: an electromagnetic-like duality invariance\footnote{Although it was first known for the photon, duality extends quite widely to other free massless fields in various dimensions, including higher $p$-forms \cite{Teitelboim:1985ya,Teitelboim:1985yc,Nepomechie:1984wu}, linearized gravity \cite{0305-4470-15-3-039,Henneaux:2004jw,Leigh:2007wf} and higher spins and representations \cite{Hull:2001iu,Bekaert:2002dt,Boulanger:2003vs,Deser:2004xt,Deser:2014ssa}.}.

Electromagnetic duality has its origins almost a century ago \cite{Dirac:1931kp,Dirac:1948um}, and has ever since been the inspiration for vast generalizations which lie at the root of many of the advances of modern theoretical physics (see e.g. the reviews \cite{Harvey:1996ur,AlvarezGaume:1997ix,Obers:1998fb}).  Electromagnetic duality is a symmetry of the sourceless Maxwell action in $3+1$ dimensional flat space \cite{Deser:1976iy,0305-4470-15-3-039}.  However, its action on the dynamical variables of the theory, the components of the gauge potential $A_\mu$, is (spatially) non-local and not manifestly Lorentz invariant.  
Acting on the gauge invariant field strength $F_{\mu\nu}\equiv \partial_\mu A_\nu-\partial_\nu A_\mu$, the duality symmetry becomes local and is manifestly Lorentz invariant, rotating the field strength into its dual: $\delta F_{\mu\nu}=\tilde F_{\mu\nu}\equiv {1\over 2}\epsilon_{\mu\nu}^{\ \ \ \lambda\sigma}F_{\lambda\sigma}$.  The equations of motion can be written in manifestly duality invariant first order form:
$d\tilde F=0,\ dF=0.$
The first equation is the field equation and the second equation is a Bianchi identity that ensures that $F_{\mu\nu}$ can be written in terms of the potential $A_\mu$.  In terms of a space+time decomposition, the anti-symmetric field strength $F_{\mu\nu}$ decomposes into the familiar electric and magnetic fields: a spatial vector $E_i=F_{i0}$ and an anti-symmetric tensor $B_{ij}=F_{ij}$ which can be dualized to a vector $B_i={1\over 2}\epsilon_{ijk}B^{jk}$.  Duality then acts to rotate the two vectors into each other: $\delta E_i=-B_i$, $\delta B_i=E_i$.  This is the form in which electromagnetic duality is traditionally known, as a symmetry of the first-order gauge invariant equations of motion for the field strength.  Duality symmetry is broken in the presence of sources, but it is in this first-order form that the introduction of magnetic sources (monopoles) becomes a natural way to restore the duality invariance.

In \cite{Deser:2013xb}, it was shown that the partially massless graviton
has a Maxwell-like duality invariance.  The duality is displayed there as a symmetry of the action in which all the gauge symmetry and auxiliary fields are stripped away.  This form of the action is advantageous for seeing clearly the physical degrees of freedom and their dynamics.  However, locality and de Sitter invariance are obscured; the action of the duality symmetry on the dynamical variables $h_{\mu\nu}$ is spatially non-local, and the de Sitter invariance is not manifest.  Here, we will display the duality in manifestly gauge invariant and de Sitter invariant form, at the level of the equations of motion.  Analogously to electromagnetic duality, it will manifest as a rotation between the first order equations of motion for the gauge invariant field strength and its Bianchi identities.  

Note that writing the equations of motion in duality covariant form cannot by itself establish duality invariance of the action.  (In fact, as in electromagnetism, the sourceless equations will be invariant under a larger GL(2) group of transformations between the field and its dual, whereas the action is only invariant under SO(2) transformations.)  The 3+1 formulation of \cite{Deser:2013xb} establishes invariance of the action, in terms of the  relevant unconstrained physical variables, at the unavoidable price of losing manifest (but of course not underlying) dS invariance.
Nevertheless, writing the equations in duality covariant form complements the 3+1 analysis and completes the analogy to electromagnetism.  In particular, it will allow us to see how to introduce magnetic sources, which cannot be introduced locally into the free action but may play a role in any eventual non-linear completion of the theory.

\section{Partially massless field equations in duality covariant form}

The field equations coming from \eqref{curvedmassivelin} are
\be \square h_{\mu\nu}-{4\over 3}\Lambda h_{\mu\nu}+\nabla_\mu\nabla_\nu h-\nabla_\mu\nabla_\lambda h_\nu ^{\ \lambda}-\nabla_\nu\nabla_\lambda h_\mu ^{\ \lambda}+g_{\mu\nu}\left(\nabla_\lambda\nabla_\sigma h^{\lambda\sigma}-\square h+{\Lambda\over 3}h\right)=0.\label{maineq}\ee
Duality is only a true symmetry of the action in the absence of a source, so we have set $T_{\mu\nu}=0$.  We comment more on sources in Section \ref{sourcesection}.

We want to show that the equations \eqref{maineq} for the potential $h_{\mu\nu}$ are equivalent to a manifestly duality invariant and gauge invariant set of equations for a field strength.  Like in electromagnetism, these should divide into field equations which reproduce \eqref{maineq}, and Bianchi identities which tell us that the field strength can be written in terms of a potential.  The duality should interchange the field equations with the Bianchi identities.

The appropriate field strength for a partially massless graviton is \cite{Deser:2006zx}
\be F_{\mu\nu\lambda}=\nabla_\mu h_{\nu\lambda}- \nabla_\nu h_{\mu\lambda},\label{fieldstrengthh}\ee
which is invariant under \eqref{gaugesym}.
Like the Maxwell field strength of the photon, it is written as a first derivative of the gauge potential and is gauge invariant.  It is anti-symmetric in the first two indices, and vanishes if all three indices are anti-symmetrized, that is, it has the symmetries of the Young tableaux {\scriptsize\Yvcentermath1 $\yng(2,1)$} in the anti-symmetric convention.

Our manifestly invariant equations will be for a field strength $F_{\mu\nu |\lambda}$ which is anti-symmetric in the first two indices but has no a-priori additional symmetries involving the third index, i.e. it is an element of the product representation {\scriptsize\Yvcentermath1 $\yng(1,1)\otimes\yng(1)$}.   
We define the dual tensor $\tilde F_{\mu\nu |\lambda}$ by dualizing over the two anti-symmetric indices using the de Sitter volume form,
\be \tilde F_{\mu\nu |\lambda}={1\over 2}\epsilon_{\mu\nu}^{\ \ \rho\sigma}  F_{\rho\sigma |\lambda}.\label{dualfdef}\ee

The equations we will find are
\bea
 && \Tr \, F=0,\ \ \ d_L \tilde F=0, \label{maineq1}\\
    && \Tr \, \tilde F=0,\ \ \ d_L  F=0. \label{maineq2}
\eea
Here, $(\Tr \ F)_\mu \equiv F^\nu_{\ \mu|\nu}$ is the only non-trivial trace, and $d_L$ is the exterior derivative with respect to the anti-symmetric pair of indices on the left, but using the full covariant derivative,
\be (d_LF)_{\mu\nu\lambda|\rho}=\nabla_\mu F_{\nu\lambda|\rho}+\nabla_\nu F_{\lambda\mu|\rho}+\nabla_\lambda F_{\mu\nu|\rho}.\label{bianchi2}\ee 
The equations \eqref{maineq1},\eqref{maineq2} are manifestly de Sitter invariant and are manifestly symmetric under the duality rotation
\be \delta F=\tilde F.\ee

In the following, we will see that the two equations \eqref{maineq2} are Bianchi identities\footnote{The Bianchi identities also make an appearance in the frame-like formulation of partially massless gravity \cite{Skvortsov:2006at}.} which tell us that the field strength is given as in \eqref{fieldstrengthh}, after which the two equations \eqref{maineq1} are field equations equivalent to \eqref{maineq}.  

\subsection{Bianchi identities}

We start with the first of the Bianchi identities \eqref{maineq2}: $\Tr\, \tilde F=0$.  In components, this reads
$ \epsilon_{\mu}^{\ \ \rho\sigma\nu}  F_{\rho\sigma |\nu}=0$.
Stripping off the epsilon, this tells us that the totally anti-symmetric part of the field strength vanishes, 
\be F_{[\mu\nu |\lambda]}=0.\label{youngs}\ee
The representation of $F_{\mu\nu|\lambda}$ can be decomposed as 
{\scriptsize\Yvcentermath1 $\yng(1,1)\otimes\yng(1)=\yng(2,1)\oplus\yng(1,1,1)$}, and \eqref{youngs} tells us that only the part with the symmetry of {\scriptsize\Yvcentermath1 $\yng(2,1)$} survives. 

Now consider the second Bianchi identity $ d_LF=0$, written out in \eqref{bianchi2}.  Once $F$ has the symmetry of {\scriptsize\Yvcentermath1 $\yng(2,1)$}, we can check explicitly that $dF_{\mu\nu\lambda\rho}\equiv\nabla_\mu F_{\nu\lambda\rho}+\nabla_\nu F_{\lambda\mu\rho}+\nabla_\lambda F_{\mu\nu\rho}$ has the symmetry of {\scriptsize\Yvcentermath1 $\yng(2,1,1)$}.  Thus we can interpret $d_L$ as a derivative operator {\scriptsize\Yvcentermath1 $\yng(2,1)\overset{d}{\longrightarrow} \yng(2,1,1)$}.  What we want is for this operator to be part of a complex with $d^2=0$ whose cohomology is trivial, so that $dF=0$ implies that $F$ can be written in terms of the gauge potential as in \eqref{fieldstrengthh}.

The description of the partially massless field involves a scalar gauge parameter $\phi$, a potential $h_{\mu\nu}$ with the symmetry {\scriptsize \yng(2)}, and a field strength $F_{\mu\nu\lambda}$ with the symmetry {\scriptsize \yng(2,1)}.

%

This, along with the desired Bianchi identity, leads us to the desired complex
 \be  \bullet\overset{d}{\longrightarrow} \yng(2)\overset{d}{\longrightarrow} \Yvcentermath1\yng(2,1)\overset{d}{\longrightarrow} \yng(2,1,1)\overset{d}{\longrightarrow}\cdots\, ,\ee 
where
\bea (d\phi)_{\mu\nu}&=&\nabla_\mu\nabla_\nu\phi+{\Lambda\over 3}g_{\mu\nu}\phi\, ,\\
(dh)_{\mu\nu\lambda}&=& \nabla_\mu h_{\nu\lambda}- \nabla_\nu h_{\mu\lambda}\, ,\\
(dF)_{\mu\nu\lambda\rho}&=&\nabla_\mu F_{\nu\lambda\rho}+\nabla_\nu F_{\lambda\mu\rho}+\nabla_\lambda F_{\mu\nu\rho}\, , \\
&\vdots& \nn
\eea
It is straightforward to check that with these definitions we have
\be d^2=0.\ee
Though we have no formal proof, on physical grounds the cohomology of this complex should be trivial given reasonable conditions on the fields\footnote{Proving this should be possible by extending the generalized Poincar\'e lemmas of \cite{DuboisViolette:2001jk,Bekaert:2002dt} to dS space.}.  In particular, if the field strength vanishes, then the potential should be pure gauge,
\be dh=0 \Leftrightarrow h=d\phi,\ee
and if a given three index hook-tableaux tensor $F$ satisfies the Bianchi identity $dF=0$, then it should be writable in terms of a potential as in \eqref{fieldstrengthh},
\be dF=0 \Leftrightarrow F=dh.\ee

\subsection{Field equations}

We now move on to the field equations \eqref{maineq2}: $ \Tr\, F=0$ and $d_L \tilde F=0$.  We want to show that these are equivalent to \eqref{maineq} once the field strength is given as in \eqref{fieldstrengthh}.  Taking a divergence of \eqref{maineq}, all the three-derivative terms cancel out and we find
\be \nabla_\nu h_\mu^{\ \nu}-\nabla_\mu h=0,\label{meq1}\ee
which is nothing but the statement that the trace of the field strength \eqref{fieldstrengthh} vanishes,
\be \Tr\, F=0.\ee
Using this to eliminate divergences in \eqref{maineq}, we have
\be \square h_{\mu\nu}-{4\over 3}\Lambda h_{\mu\nu}-\nabla_\mu\nabla_\nu h+{\Lambda\over 3}hg_{\mu\nu}=0.\label{meq2}\ee
Now consider the expression,
\be \nabla^\lambda F_{\lambda \mu\nu}\sim \epsilon^{\mu_1\mu_2\mu_3}_{\ \ \ \ \ \ \ \mu}\nabla_{\mu_1}\left(\epsilon_{\mu_2\mu_3 \nu_1\nu_2}F^{\nu_1\nu_2}_{\ \ \ \ \ \nu}\right)=0.\label{meq3}\ee
Expanding \eqref{meq3} using \eqref{fieldstrengthh} and then using \eqref{meq1} to eliminate divergences, this reproduces \eqref{meq2}.
Stripping off the first epsilon, \eqref{meq3} gives $\nabla_{[\mu_1}\left(\epsilon_{\mu_2\mu_3] \nu_1\nu_2}F^{\nu_1\nu_2}_{\ \ \ \ \ \nu}\right)=0$, which is nothing but
\be d_L \tilde F=0.\ee

Note that the action \eqref{curvedmassivelin} with the partially massless tuning \eqref{masstuning} is gauge invariant under \eqref{gaugesym} for any Einstein space background with cosmological constant $\Lambda$.  However, the duality results of \cite{Deser:2013xb} and those presented here go through only if the background is de Sitter.  Indeed, even the field strength \eqref{fieldstrengthh} fails to be gauge invariant for an Einstein space which is not maximally symmetric, instead picking up a piece proportional to the background Weyl tensor.  This is reminiscent of the results of \cite{Deser:1997gq,Bunster:2012hm,Moon:2014gaa} linking duality invariance and maximal symmetry.

\section{$3+1$ decomposition and Maxwell-like equations}

In this section we perform a space+time decomposition and display the equations of motion and duality in a form analogous to the traditional undergraduate presentation of Maxwell's equations.  In this form, it will be easy to see why $3+1$ dimensions is special, because this is the only dimension in which the spatial tensors arrange into pairs which can rotate into each other under duality.

We write the background de Sitter metric in the flat inflationary coordinates,
\be ds^2=-dt^2+a(t)^2d\vec x^2,\ \ \ a(t)=e^{Ht},\ee
where $H=\sqrt{\Lambda\over 3}$ is the de Sitter Hubble constant.  

We will work with the field strength $F_{\mu\nu\lambda}$ for which the two algebraic equations $\Tr \, F=0$ and $\Tr\, \tilde F=0$ have already been solved, so that the field strength is in the representation {\scriptsize\Yvcentermath1 $\yng(2,1)^{\ T}$}  (the superscript $T$ indicates that the tensor is traceless).  Consider for a moment this tensor in $d+1$ spacetime dimensions.  As an (irreducible) representation of $so(d+1)$, it decomposes upon restriction to $so(d)$ as,
\be {\Yvcentermath1 \yng(2,1)}^{\ T}\underset{so(d+1)\rightarrow so(d)}{\longrightarrow} \Yvcentermath1 \yng(2,1)^{\ T}\oplus \yng(2)^{\ T}\oplus \yng(1,1) \oplus \yng(1)\ , \ee
and this decomposition can be implemented as follows:
\bea F_{\mu\nu\lambda}=\begin{cases} F_{i00}=a E_i \\  F_{i0j}=a^2\left(H_{ij}+{1\over 2}B_{ij}\right) \\ F_{ij0}=a^2 B_{ij} \\ F_{ijk}= a^3 \left[ f_{ijk}+{1\over d-1}\left(E_i\delta_{jk}-E_j\delta_{ik}\right)\right] .\end{cases}
\eea
Here $E_i={\scriptsize\Yvcentermath1 \yng(1)^{\ }}$ is a spatial vector , $H_{ij}={\scriptsize\Yvcentermath1 \yng(2)^{\ T}}$ a symmetric traceless tensor, $B_{ij}={\scriptsize\Yvcentermath1 \yng(1,1)^{\ }}$ an anti-symmetric tensor and $f_{ijk}={\scriptsize\Yvcentermath1 \yng(2,1)^{\ T}}$ a traceless mixed-symmetry tensor. 

 For $d=3$, we can dualize the antisymmetric tensor $B_{ij}$ into a pseudo-vector $B_i$ and the mixed-symmetry tensor $f_{ijk}$ into a symmetric traceless pseudo-tensor $K_{ij}$,\footnote{Spatial indices are always moved with $\delta_{ij}$, and $\epsilon_{ijk}$ is the standard flat space epsilon symbol with $\epsilon_{123}=1$.}
 \be B_i={1\over 2}\epsilon_{ijk}B^{jk},\ \ \ \ \ K_{ij}={1\over 2}\epsilon_{ikl}f^{kl}_{\ \ j}.\ee
 (Tracelessness of $K_{ij}$ follows from the fact that $f_{ijk}$ has no totally anti-symmetric component, and symmetry of $K_{ij}$ follows from tracelessness of $f_{ijk}$.)  Thus, only in $d=3$, the gauge-invariant spatial fields arrange into a pair of vectors $B_i,E_i$, and a pair of symmetric traceless tensors $H_{ij},K_{ij}$.  

In terms of these spatial variables the field equations take the following form in vector calculus notation\footnote{Starting from the field equations as in \eqref{meq3}, the 00 equation gives \eqref{fieldeq1}, the $0i$ equation gives \eqref{fieldeq2}, the $i0$ equation gives \eqref{fieldeq3}, and the symmetric traceless part of the $ij$ equation gives \eqref{fieldeq4}.  The anti-symmetric part of the $ij$ equations is redundant with the Bianchi equations \eqref{bianchieq2} and \eqref{bianchieq3}, and the trace of the $ij$ equations is redundant with \eqref{fieldeq1}.},
\bea && \nabla\cdot \vec E=0\, , \label{fieldeq1}\\
&& \nabla\cdot \overset{\leftrightarrow}{H}-{1\over 2}\nabla\times \vec B-Ha\vec E=0\, , \label{fieldeq2}\\
&& \left({d\over dt}+3H\right)\vec E-{1\over a}\nabla\times\vec B=0\, ,\label{fieldeq3}\\
&&  \left({d\over dt}+2H\right)\overset{\leftrightarrow}{H}-{1\over a}\nabla\odot \vec E-{1\over a}\nabla\times \overset{\leftrightarrow}{K}=0\, ,
\label{fieldeq4}
\eea
where $\left(\nabla\odot \vec E\right)_{ij}\equiv{1\over 2}\left( \partial_iE_j+\partial_jE_i-{1\over 3}\delta_{ij}\nabla\cdot \vec E\right)$ is the symmetrized traceless derivative and $\left(\nabla\times \overset{\leftrightarrow}{H}\right)_{ij}\equiv {1\over 2}\left( \epsilon_{ikl}\partial^k H^l_{\ j}+\epsilon_{jkl}\partial^k H^l_{\ i}\right)$ is the symmetrized curl. 
The Bianchi identities take the form\footnote{Writing the Bianchi identity as $\epsilon_{\mu}^{\ \mu_1\mu_2\mu_3}\nabla_{\mu_1}F_{\mu_2\mu_3\nu}=0$, the 00 equation gives \eqref{bianchieq1}, the $0i$ equation gives \eqref{bianchieq2}, the $i0$ equation gives \eqref{bianchieq3}, and the symmetric traceless part of the $ij$ equation gives \eqref{bianchieq4}.  The anti-symmetric part of the $ij$ equations is redundant with the field equations \eqref{fieldeq2} and \eqref{fieldeq3}, and the trace of the $ij$ equations is redundant with \eqref{bianchieq1}.
},
\bea && \nabla\cdot \vec B=0\, , \label{bianchieq1}\\
&& \nabla\cdot \overset{\leftrightarrow}{K}+{1\over 2}\nabla\times \vec E-Ha\vec B=0\, , \label{bianchieq2}\\
&& \left({d\over dt}+3H\right)\vec B+{1\over a}\nabla\times\vec E=0\, ,\label{bianchieq3}\\ 
&&  \left({d\over dt}+2H\right)\overset{\leftrightarrow}{K}-{1\over a}\nabla\odot \vec B+{1\over a}\nabla\times \overset{\leftrightarrow}{H}=0.
\label{bianchieq4}
\eea

From these first order equations we can easily verify the number of propagating degrees of freedom: the vectors each have 3 components and the traceless symmetric tensors each have 5 components, and they obey the first order (in time) equations \eqref{fieldeq3}, \eqref{fieldeq4}, \eqref{bianchieq3}, \eqref{bianchieq4} for a total of 16 initial conditions.  These must obey the 8 constraint equations \eqref{fieldeq1}, \eqref{fieldeq2}, \eqref{bianchieq1}, \eqref{bianchieq2}, which brings the number of independent initial data down to 8.  These are the configurations and conjugate momenta for the 4 degrees of freedom of the partially massless graviton in $3+1$ dimensions.  Roughly, the helicity one mode is captured by the spatial vectors and the helicity two mode is captured by the spatial symmetric tensors.  Note that \eqref{fieldeq1}, \eqref{fieldeq3},  \eqref{bianchieq1}, \eqref{bianchieq3} are autonomous equations for $\vec E$ and $\vec B$ which are nothing but Maxwell's equations on de Sitter space.

 Defining analogous spatial tensors $\tilde B_i$, $\tilde E_i$, $\tilde H_{ij}$, $\tilde K_{ij}$ for the dual tensor $\tilde F_{\mu\nu\lambda}$, we find after expanding \eqref{dualfdef} that the effect of duality is to rotate the two vectors into each other and the two tensors into each other,
\bea && \tilde E_i=-B_i,\ \ \ \ \tilde B_i=E_i,\\ 
&& \tilde H_{ij}=-K_{ij},\ \ \ \ \tilde K_{ij}=H_{ij}.
\eea
The Maxwell-like equations are manifestly invariant under duality; the field equations are rotated into the Bianchi identities.

\section{Sources\label{sourcesection}}

In the case of electromagnetism, the presence of a source $j_\mu$ alters the right-hand side of the field equations: $d\tilde F=\ast j$, $dF=0$.  This breaks the duality symmetry, but we may restore it by introducing a magnetic source $\tilde j_\mu$ which appears on the right-hand side of the Bianchi identity and predicts the existence of magnetic monopoles: $d\tilde F=\ast j$, $dF=\ast \tilde j$.  Magnetic sources cannot be introduced locally into the free action, but play a crucial role in non-linear embeddings of the theory \cite{'tHooft:1974qc,Polyakov:1974ek}.

Consider restoring the source $T_{\mu\nu}$ into the partially massless theory.  As with electromagnetism, the source makes an appearance on the right-hand side of the field equations \eqref{maineq1},
\bea
 && \Tr \, F={3\over 2\Lambda}\nabla T,\ \ \ d_L \tilde F=\ast'\, T, \label{maineq1s}\\
    && \Tr \, \tilde F=0,\ \ \ d_L  F=0, \label{maineq2s}
\eea
where $(\ast' \,T)_{\mu_1\mu_2\mu_3\mu_4}\equiv T_{\mu_4}^{\ \nu}\epsilon_{\nu\mu_1\mu_2\mu_3}+{1\over 2}T^\nu_{\ \nu}\epsilon_{\mu_1\mu_2\mu_3\mu_4}-{3\over 2\Lambda}\nabla_{\mu_4}\nabla_\nu T^{ \nu\lambda}\epsilon_{\lambda\mu_1\mu_2\mu_3}$.  Duality symmetry is broken by the source.

This invites us to introduce a ``magnetic" source tensor $\tilde T_{\mu\nu}$ that satisfies the same conservation equation \eqref{Tconservat} as the original source tensor, and which acts as a source for the Bianchi identities and restores duality invariance to the equations,
\bea
 && \Tr \, F={3\over 2\Lambda}\nabla T,\ \ \ d_L \tilde F=\ast'\, T , \label{maineq1ss}\\
    && \Tr \, \tilde F={3\over 2\Lambda}\nabla \tilde T,\ \ \ d_L  F=\ast'\, \tilde T. \label{maineq2ss}
\eea
Note that in the presence of generic magnetic sources, the field strength will have an anti-symmetric component and will not be in a pure mixed symmetry representation.  The interpretation of such a dual source may be something along the lines of \cite{Bunster:2006rt,Bunster:2013era}.  As with electromagnetism, such a dual source cannot be introduced locally into the free action, but may play a role in any eventual non-linear embedding of the theory.

{\bf Acknowledgements:} The author would like to thank Andrew Waldron for many helpful discussions and comments.  Research at Perimeter Institute is supported by the Government of Canada through Industry Canada and by the Province of Ontario through the Ministry of Economic Development and Innovation.  This work was made possible in part through the support of a grant from the John Templeton Foundation. The opinions expressed in this publication are those of the author and do not necessarily reflect the views of the John Templeton Foundation.

 \bibliographystyle{utphys}
\addcontentsline{toc}{section}{References}
\bibliography{PMdualityarxiv}

\end{document}